\documentclass[%
reprint, 
superscriptaddress,
 amsmath,amssymb,
 aps, physrev,
]{revtex4-2}
\usepackage[english]{babel}
\usepackage{graphicx}% Include figure files
\usepackage{dcolumn}% Align table columns on decimal point
\usepackage{bm}% bold math
\usepackage{hyperref}
\usepackage{soul}% add hypertext capabilities
\usepackage{xcolor}
\usepackage{adjustbox}
\usepackage{orcidlink}
\usepackage{textcomp}
\DeclareUnicodeCharacter{223C}{\textasciitilde}
\begin{document}

%\preprint{APS/123-QED}

\title{\textbf{Trapping of nanofabricated SiO$_2$ high-aspect-ratio discs in an optical lattice} 
}% 

\author{Shelby Klomp\,\orcidlink{0000-0002-5204-7455}}
\altaffiliation{These authors contributed equally to this work}
\affiliation{%
Center for Fundamental Physics, Department of Physics and Astronomy, Northwestern University, Evanston, Illinois 60208, USA
}%

\affiliation{Center for Interdisciplinary Exploration and Research in Astrophysics (CIERA), Northwestern University, Evanston IL 60208, USA.}

\author{Shafaq Gulzar Elahi\,\orcidlink{0000-0002-3671-692X}}
\altaffiliation{These authors contributed equally to this work}
\affiliation{%
Center for Fundamental Physics, Department of Physics and Astronomy, Northwestern University, Evanston, Illinois 60208, USA
}%
\affiliation{Center for Interdisciplinary Exploration and Research in Astrophysics (CIERA), Northwestern University, Evanston IL 60208, USA.}
\author{Scott Grudichak}
\affiliation{%
Center for Fundamental Physics, Department of Physics and Astronomy, Northwestern University, Evanston, Illinois 60208, USA
}%
\author{Chethn Galla}
\author{Zhiyuan Wang}
\author{Mark Nguyen}
\author{George Winstone}
\author{Katarina Boskovic Guy}
\author{Emma Derrick}
\affiliation{%
Center for Fundamental Physics, Department of Physics and Astronomy, Northwestern University, Evanston, Illinois 60208, USA
}%
\author{Maddox Wroblewski}
\author{Jacob Sprague}
\affiliation{%
Center for Fundamental Physics, Department of Physics and Astronomy, Northwestern University, Evanston, Illinois 60208, USA
}%
\affiliation{Center for Interdisciplinary Exploration and Research in Astrophysics (CIERA), Northwestern University, Evanston IL 60208, USA.}

\author{Andrew Laeuger}
\affiliation{Department of Physics, Caltech, CA, USA}
\author{Jackson Larsen}
\author{Nancy Aggarwal\,\orcidlink{0000-0002-0026-3877}}
\affiliation{Department of Physics, University of California at Davis, Davis, CA, USA}
\author{Shane L. Larson}
\affiliation{Department of Physics and Astronomy, Clarkson University, USA}
\author{Andrew A. Geraci \,\orcidlink{0000-0001-7009-0118}}%
\email{andrew.geraci@northwestern.edu}
\affiliation{%
Center for Fundamental Physics, Department of Physics and Astronomy, Northwestern University, Evanston, Illinois 60208, USA
}%
\affiliation{Center for Interdisciplinary Exploration and Research in Astrophysics (CIERA), Northwestern University, Evanston IL 60208, USA.}
\collaboration{LSD Collaboration}%\noaffiliation

\date{\today}% It is always \today, today,

\begin{abstract}
We describe the nanofabrication and loading of lithographically patterned, high aspect ratio $10$ $\mu$m diameter silicon dioxide discs with thickness of $400$ nm into an optical standing wave trap in vacuum. We obtain motional spectra of the center of mass motion of the discs using a combination of interferometric and imaging methods. The method is capable of high-efficiency loading and optical trapping using small numbers of objects, of the order of a few particles, with customized geometries and sizes compatible with wafer scale nanofabrication methods. Disc-like particles in optical traps have been identified as promising candidates for detectors of high frequency gravitational waves or as sensitive accelerometers.   

\end{abstract}

\maketitle

%\tableofcontents

\emph{Introduction.}
%\textcolor{red}{This is a very rough first pass at an intro, feel free to edit}
%\textcolor{red}{Planning to target PRA as a journal, also considering Optics Letters or APL Photonics, but if we get stacks then maybe PRL}
%\textcolor{red}{Add authors: Mark, Emma, Jacob, Yuqi?}
%\textcolor{red}{To do list:
%+ in intro: point out first time round disc trapped in vacuum, add James Millen rod paper,
%+ figure 1: zoom in on standing wave, remove launcher picture, 
%+ new figures: (2) new experiment diagram (add more SEMS), more data, (3) simulation results
%\textcolor{red}{
%+ Andy: intro, GW sensitivity, conclusion, paragraph about pendulum motion,
%+ Shelby: fig1, fig2 (more SEMs),fig3
%+ Shafaq:  fig4 (damping), dynamics section,
%+ Max: damping analysis,
%+ Scott: longer fabrication appendix
%}
Levitated optomechanics \cite{levreview,Millen:2020review,levitodynamics_review}, involving the optical measurement and manipulation of objects suspended with electromagnetic forces such as radiation pressure, electrostatic trapping, ion traps or magnetic interactions, has emerged over the past decade as a rich area of research with many applications including precision measurement of forces \cite{Geraci2010,ranjit_zeptonewton_2016, montoya_scanning_2022}, accelerations \cite{andyhart2015, Moore2017,novotnydrop,monteiro_force_2020}, torques \cite{Li2016} or rotations \cite{Li2018,Novotny2018,Moore2018}, or electromagnetic fields \cite{blakemore_three-dimensional_2019,Moore_Millicharged_2014}. Such systems have been proposed for tests of fundamental physics (for a recent review see Ref. \cite{moore2021searching}), including searches for new exotic interactions \cite{Geraci2010}, dark matter \cite{tseng_search_2025,Hamaide2026}, nuclear decays \cite{mooredecay}, and gravitational waves \cite{Arvanitaki:2013,aggarwal2022searching, winstone_optical_2022,FormalismPaper,DielectricLetter,AndrewsGRNote}.
%\textcolor{red}{[add Laeuger citation?]}. 
Recently the ability to control such systems in the quantum regime, including ground state cooling \cite{tebbenjohanns_quantum_2021, delic_cooling_2020, piotrowski_simultaneous_2023, ranfagni_two-dimensional_2022}, squeezing \cite{kamba_quantum_2025}, delocalization, \cite{kremer2026fastquantumsqueezingnanomechanical,rossi2024quantumdelocalizationlevitatednanoparticle} and sensing below the standard quantum limit \cite{rossisubsql} has opened avenues for investigations of foundational aspects of quantum mechanics \cite{Bateman2014,ORI11_GM,Goldman2015} including tests of quantum collapse models \cite{ORI11_GM,Vinante_2019} as well as uses in quantum information science and quantum communication \cite{Mancini2003, deplano2026stationaryentanglementlevitatedoscillator}.

Thus far, the vast majority of work in levitated optomechanics has involved using small spherical particles of sub-wavelength size which are amenable to trapping in a single-beam optical tweezer or spherical particles of sizes ranging up to tens of micrometers which can be confined in counter-propagating dual beam traps \cite{li_millikelvin_2011,ranjit_attonewton_2015} or vertically oriented optical levitation traps \cite{Ashkin:1971,Ashkin:1976,Ashkin:1977,Moore2017,Moore_Millicharged_2014}. 
 %By expanding the possibilities of shapes, sizes, and materials of the levitated object, several other applications become possible.
%The field of levitated optomechanics has shown great promise for applications including...
%These applications may require a wide range of shapes and sizes of particles to be trapped. 
There are a number of applications in levitated optomechanics that benefit from or require the use of non-spherical shaped particles. For instance, torque and rotation sensing \cite{Li2016, Li2018,Novotny2018} has been explored using nanodumbells. % or gravitational wave detection\cite{}, using high aspect ratio hexagonal NaYF plate-like prisms. \
For gravitational wave detection, it is desirable to trap high mass objects at high frequencies which can be achieved using flat plate-like particles in an optical standing wave trap \cite{aggarwal2022searching}, for example high aspect ratio plate-like NaYF hexagonal prisms \cite{winstone_optical_2022}. Such geometries also present advantages for reducing heating due to photon recoil \cite{aggarwal2022searching}.
In addition, protocols that have been proposed for experimentally verifying whether gravity can mediate quantum entanglement between massive objects \cite{bose_spin_2017,marletto_gravitationally_2017} may benefit from particles of customized geometry, e.g. cylindrical shaped nanodiamonds with precisely oriented and positioned embedded NV- centers, where the cylindrical geometry can constrain their equilibrium orientation in certain trapping potentials \cite{elahi_diamagnetic_2025}.

Levitated optomechanics experiments that suspend dielectric objects in vacuum by using radiation pressure have primarily used silica nanospheres that can be fabricated in large quantities.
Some of the most common loading techniques for delivering nano- to micro-sized particles into levitated systems include nebulizers \cite{AspelmeyerCooling,Novotny2012} and piezoelectric transducers \cite{weisman2022,khodaee2022,li_millikelvin_2011}, which typically require large initial quantities of particles in order to successfully load particles into the trap, although larger ($\sim 10$ $\mu$m) silica microspheres have been captured, released, and re-captured on-demand by shaking a transparent substrate holding the particles in a vertically oriented optical levitation trap \cite{monteiro_force_2020}.

In many cases these particles used in optical levitation experiments are grown from a bottom-up approach.
For example silica spheres grown via the Stober process are routinely used but suffer from density inhomogeneities and dispersion in their size, typically at the $\sim 10$ percent level.  Depending on the surfactant used in synthesis, $\beta$-phase NaYF can be fabricated into flat plate like hexagonal prisms using a bottom-up fabrication method \cite{winstone_optical_2022}. However these particles tend to have size variation on the order of $\sim$ 10-20 percent within a single batch. 

%Lithographic techniques 
\textcolor{black}{Top-down fabrication techniques such as lithography, where the particles are fabricated from bulk material,} offer more control over the shape and dimensions of the dielectric object to be suspended. Recent work has shown that large quantities of lithographically produced silicon cylinders \cite{siliconnew} and nanorods \cite{Millen_nanorod} can be loaded into optical traps, where silicon has a higher refractive index than silica, making the engineering of optical forces through Mie resonances more feasible.  
%Existing techniques include nebulizers \cite{AspelmeyerCooling,Novotny2012}, piezoelectric transducers \cite{weisman2022,khodaee2022,Li:2011}, laser-induced acoustic desorption (LIAD) \cite{MillenLIAD,northup}, transferring a particle to ultra-high vacuum in a hollow-core fiber  \cite{HCFload}, and particle release by sublimation of camphor \cite{dursocamphor}. 
 Generally, lithographic techniques allow for the fabrication of $\sim \mu$m-sized objects that can be made very uniform, but often only produce smaller quantities than bottom-up synthesis, making them more challenging to trap using loading techniques which rely on large quantities of particles. 
 %\textcolor{violet}{(Here, you don't explicitly say that lithographic techniques are top-down. Also, I didn't really know what top-down vs. bottom-up meant until I looked it up. -ED)}

%In a recent work, we developed an apparatus that allows for more localized efficient loading using a pulled glass capillary micropipette\cite{Grinin2026}. This apparatus allowed for loading of a range of sizes and shapes of particles, and demonstrated trapping with as low as ~100,000 particles.

%In this Letter, we demonstrate in-vacuum optical trapping of high-aspect-ratio disc-shaped SiO$_2$ particles with size determined by nanofabrication methods. 
%\textcolor{blue}{From Nancy: 
In this Letter, we demonstrate high-efficiency in-vacuum optical trapping of lithographically nanofabricated high-aspect-ratio disc-shaped SiO2 particles.
The loading method we employ, using our developed micropipette-based launching methods reported in Ref. \cite{Grinin2026}, requires only a few ($<10$) particles in order to obtain a successful trapping event. This work establishes the experimental feasibility of loading optical traps with small quantities of dielectric objects with dimensions of order $\sim 10$ $\mu$m that are produced by wafer-scale top-down nanofabrication techniques. The high aspect ratio disc geometry we study in particular may be useful for levitated-optomechanics-based searches for high frequency gravitational waves. 

%\section{Experimental Methods}

\begin{figure}
    \includegraphics[width=0.5\textwidth]{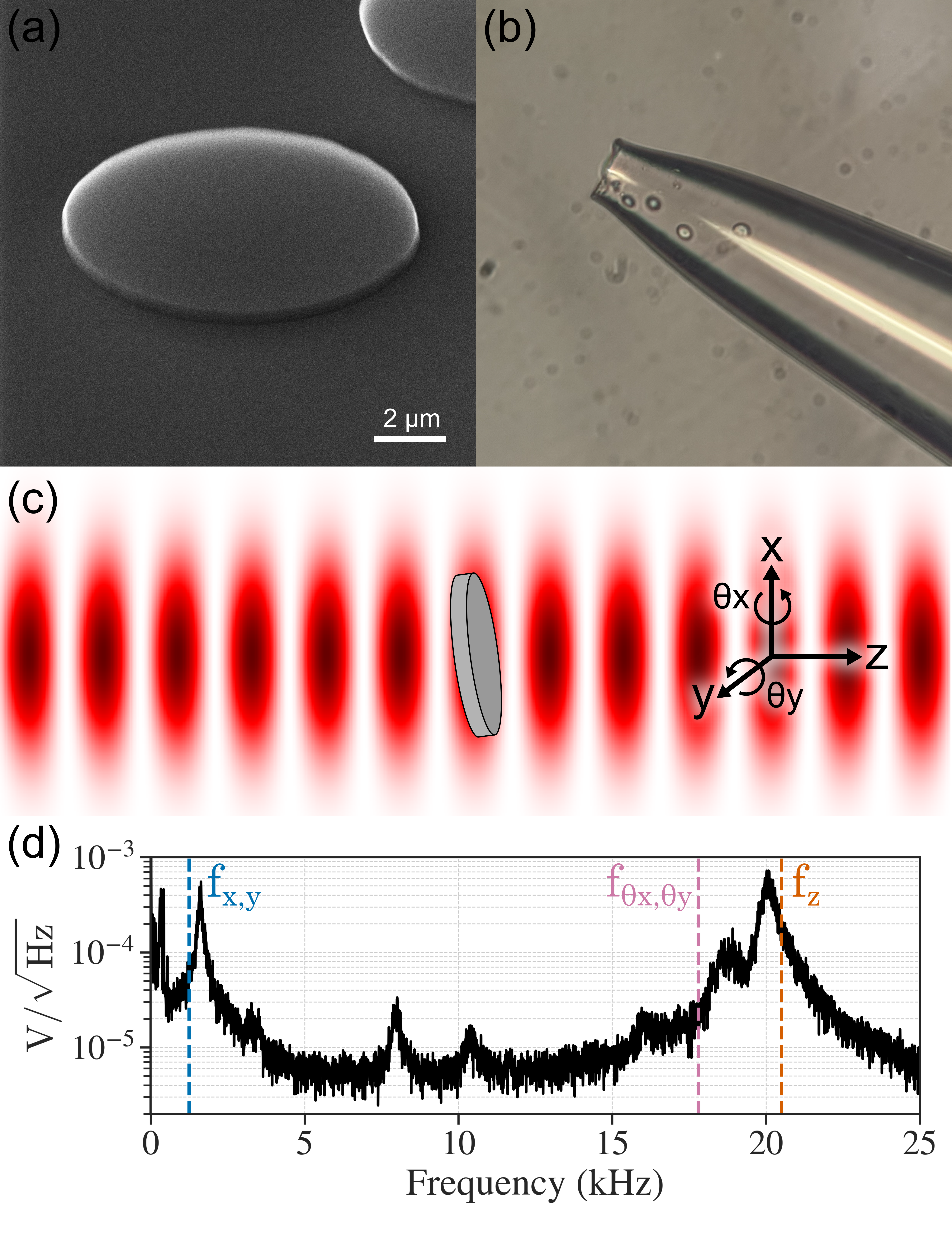}
    \caption{(a) SEM image of SiO$_2$ disc with 10$~\mu$m diameter and $400$ nm thickness. Viewing angle is 60 degrees from the surface normal, (b) optical microscope photo of 4 discs in the tip of a micropipette with a 30 $\mu$m opening, (c) depiction of a disc in a standing wave optical trap with motional coordinate system, (d) amplitude spectral density of disc in trap recorded with a fiber balanced photodetector positioned in the z direction. The dashed vertical lines correspond to predicted frequencies for the motional peaks corresponding to the transverse (x,y), librational ($\theta_x,\theta_y$), and axial (z) motions.}
    \label{fig:fig1}
\end{figure}
% bigger font on (d) and bigger coordinate system
% label peaks
\emph{Particle Fabrication.}
% The disc-shaped SiO$_2$ particles are fabricated using high-pressure plasma-enhanced chemical vapor deposition, etc...
The disc-shaped SiO$_2$ particles were fabricated using a top-down lithographic process 
%\textcolor{violet}{(This is the first place you say that the particles are fabricated using a top-down approach. -ED)} 
also described in Ref.~\cite{Galla2023}. The process involved coating a 100 mm diameter, $500~\mu$m thick Si wafer with a protective low-stress Si$_3$N$_4$ layer through a low pressure chemical vapor deposition (CVD) process to serve as an etch stop. A $1~\mu$m amorphous-silicon sacrificial layer, and a 400~nm SiO$_2$ device layer were then
both deposited with a high density plasma CVD process. The SiO$_2$ layer was patterned by optical lithography using a $1~\mu$m layer of a positive photoresist and exposed with light to protect the top of the discs while leaving the space around the discs soluble to the chemical developer. After development, the patterned substrate underwent an inductively coupled anistotropic plasma fluorine etch to clear 
%\textcolor{violet}{(Typo: "clearing" should be "clear". -ED)} 
the space around the discs. A complete clearing of the SiO$_2$ surrounding the discs is necessary for the isotropic wet etch of the a-Si layer.
The discs were released by selectively etching the a-Si layer with a heated 25\% solution of Tetramethylammonium hydroxide (TMAH).
Using these methods, we have fabricated over $10^6$ free-standing silicon dioxide discs per wafer with a measured diameter of $10~\pm~0.5~\mu$m and thickness of $400~\pm~10$ nm for the batch shown in Fig. \ref{fig:fig1}.
%\textcolor{red}{(this is for Mark's batch but later we say the size of the old batch and they don't agree)}. 
A scanning electron microscope (SEM) image of one of these discs is shown in Fig. \ref{fig:fig1}(a). 

%\textcolor{purple}{MN: The low stress nitride layer was grown on both sides of a silicon substrate through a low pressure chemical vapor deposition process (CVD). Both the 1$\mu m$ layer of a-Si as well as the 400$nm$ device layer of SiO$_2$ were both deposited with a high density plasma CVD process. The substrate was then spin coated with a ~1$\mu m$ layer of a positive photoresist and exposed with light to protect the top of the discs while leaving the space around the discs soluble to the chemical developer. After chemical development, the patterned substrate underwent an inductively coupled plasma Fluorine etch, with the intention of anisotropically clearing the spacing around the discs. A complete clearing of the SiO$_2$ surrounding the discs is necessary for the isotropic wet etch of the a-Si layer. Following the wet etch, the solution containing the released discs is diluted many times with deionized(DI) water to achieve parts per hundred million(pphm) of TMAH for safe handling. This pphm solution is evaporated on a hot plate, to create a higher concentration of discs for ease of loading as outlined in the following section. The Si$_3$N$_4$ substrate that remains is carefully extracted and rinsed with DI water as there are still some discs that have released but are adhered through stiction forces.}

\begin{figure}
    \includegraphics[width=0.5\textwidth]{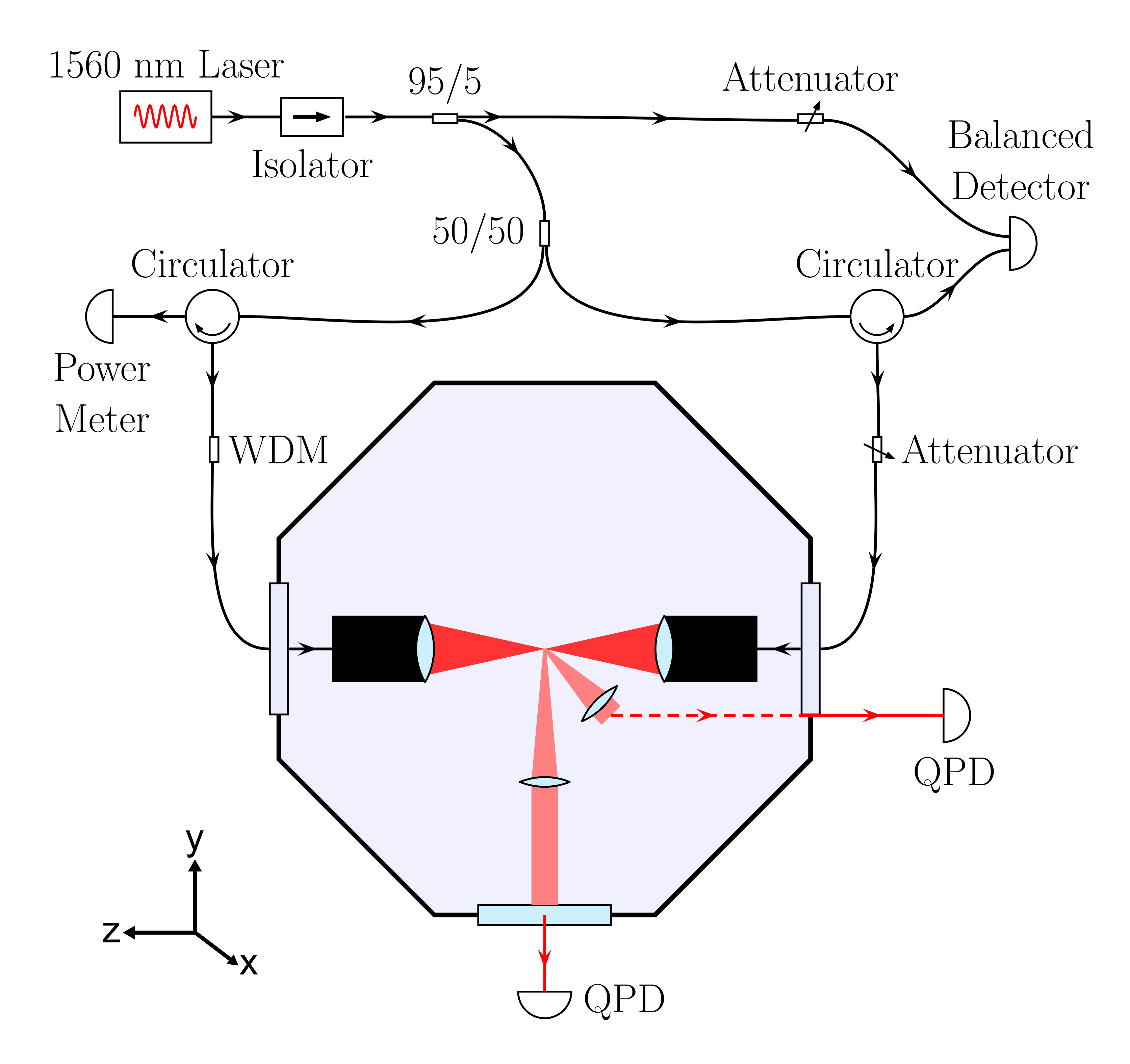}
    \caption{Diagram of the optical trapping apparatus consisting of a 1560 nm laser forming a dual-beam standing wave optical trap in a vacuum chamber. The motion of a disc is recorded by imaging the scattered trap light onto segmented quadrant photodiodes (QPD)s or by using a homodyne-like interference signal collected through the fiber onto a balanced detector \cite{winstone_optical_2022}. A wave division multiplexer (WDM) allows the introduction of light at a separate wavelength which can be used for linear optical feedback cooling.}
    \label{fig:fig2} 
\end{figure}

%\emph{Optical trap loading.}
\emph{\textcolor{black}{Optical trapping methods.}}
The optical trapping apparatus is a dual-beam counter-propagating standing wave trap using a laser of wavelength $1560$ nm. Each of the beams is linearly polarized and focused to a beam waist of approximately $5$ $\mu$m at the trapping region. A diagram of the apparatus is shown in Fig. \ref{fig:fig2}. Further detail of the trapping apparatus has been described in Ref. \cite{winstone_optical_2022}.

The particles are delivered into the optical trap using a piezo-based micropipette launcher as described in Ref. \cite{Grinin2026}. In this case, the $10$ $\mu$m discs are loaded directly into the $30$ $\mu$m opening of the micropipette tip
%using the ``anteater" method depicted in Fig. \ref{fig:fig1}(b). In this method, we 
by preparing a microscope slide of discs and mounting the empty micropipette launcher to a micromanipulator next to an optical microscope where the micropipette is at an angle approximately $35-40$ degrees up from the surface. %Then under the microscope, the micropipette tip is touched down gently to the surface of the microscope slide and carefully moved laterally along the surface to pick up discs one at a time.
%The angle of the pipette is important because at angles higher than $\sim 45$ degrees, the particles do not get picked up over the lip of the pipette opening, whereas at angles less than $\sim 30$ degrees, the tip is much more easily broken when it makes contact with the surface of the slide.
After the discs are loaded into the tip of the micropipette, the piezoelectric launcher apparatus is positioned in the chamber and aligned using 3D piezoelectric positioning stages to be $0.5-2$ mm above the trapping region of the optical trap. The particles are launched and trapped at a pressure of 6 mbar.
%can also talk about the trapping power and pressure

We have successfully trapped single particles after loading $5-10$ individual particles near to the 30 $\mu$m opening of the micropipette as shown in Fig. \ref{fig:fig1}(b). We have trapped up to two individual discs from a single loaded pipette from successive launching attempts. This has been done repeatedly for 
% a total of 8 
multiple successfully trapped discs using this method. 

The motion of the trapped particle is observed using various detection schemes as depicted in Fig. \ref{fig:fig2}. There are two collection lenses that send scattered light to quadrant photodiodes (QPD) positioned to the side and bottom of the vacuum chamber which image the motion of the particle. Additionally, light that is collected back through one of the trapping lenses travels back through a circulator to a balanced detector which observes the motion of the particle interferometrically \cite{winstone_optical_2022}.

% try putting particles bigger across the bottom

\emph{Characterization of the Motional Spectrum.}
%\subsection{Observed Dynamics}
\begin{figure*}
    \includegraphics[width=1\textwidth]{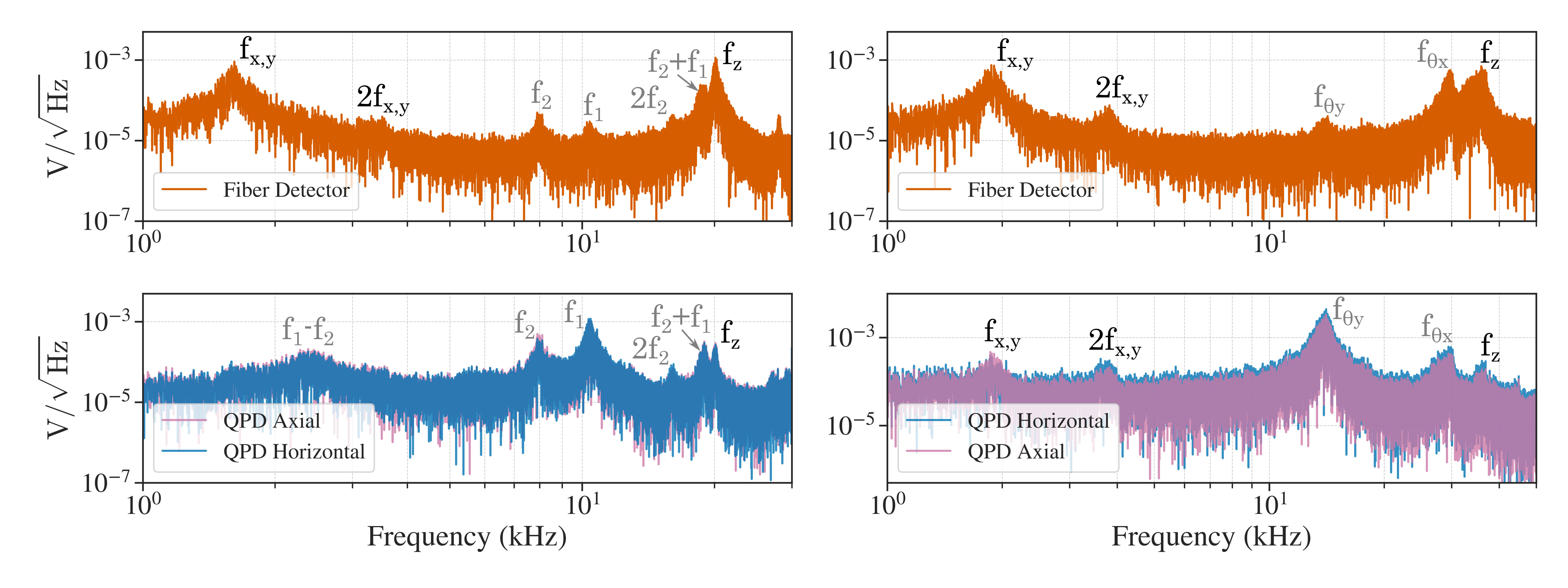}
    \caption{Motional spectra representing typical discs in the optical trap. The left plots are the same particle as Fig. \ref{fig:fig1}, \textcolor{black}{trapped in the standing wave with 50~mW power per beam} 
    %at 500 mW (set) 
    at 0.25 mbar. The right plots are an additional particle, \textcolor{black}{trapped at 60~mW power per beam} 
    %600 mW (set) 
    at 0.3 mbar. For each particle, spectra are shown for the fiber-based detector in the top plot and two channels of a QPD in the bottom plot. Prominent peaks are labeled indicating assignments of the trap normal modes as further explained in the text. We postulate that peaks labeled in gray are associated with librational modes of the disc in the presence of imperfections.}
\label{fig:fig3} 
\end{figure*}
%One example of a power spectral density (PSD) for an optically levitated disc is shown in Figure \ref{fig:fig1}(e). We do observe the expected 5 primary peaks corresponding to translational and librational motion as has been observed for similar plate-like geometries\cite{Winstone2022}.
%In addition to the translational and librational motion, we also observe additional peaks in the spectrum that we are still working to explain...
%($R/w\sim $$\mathcal{O}(1)$)
The nanofabricated silica discs we have trapped thus far were fabricated in an earlier batch than the one depicted in Fig. \ref{fig:fig1} and have a slighly smaller measured radius $r = (4.5\pm 0.5)~\mu$m and similar thickness $t = 400$ nm. For such a particle, $k r\approx 20$ where $k$ is the wavenumber
(which lies in the deep Mie regime). Therefore,  in order to understand how these particles interact with the trapping optical field, we use SCUFF-EM (Surface CUrrent/Field Formulation of ElectroMagnetism)~\cite{SCUFF1,SCUFF2}---a Boundary-Element-Method (BEM) implementation of electromagnetic scattering to obtain the optical forces and torques and subsequently the frequencies of the particle corresponding to motion along different degrees of freedom. 
%To obtain the optical force and torque and subsequently the frequencies, we use the Displaced Surface Integral Method (DSIPFT). Since SCUFF-EM is a boundary method solver, it offers a substantial speed up over other DDA solvers for micron sized object. 
%$I_{\parallel} = \frac{1}{2}MR^{2} \qquad \text{(about the symmetry axis: spin, } \phi\text{)}
%I_{\perp} = \frac{1}{12}M\left(3R^{2}+H^{2}\right) \qquad \text{(about any diameter: tilt, } \theta_x, \theta_y\text{)}$
For a plate-like particle geometry in an standing wave trap with co-linearly polarized light, we expect to observe three peaks corresponding to translational motion ($x$, $y$, and $z$) and two peaks corresponding to librational motion ($\theta_x$ and $\theta_y$). The third librational motion, ($\theta_z$) is about the symmetry axis of the disc and is not confined by the optical potential for a perfectly circular disc and beam, and in the absence of birefringence. The definitions of these motional directions with respect to the trap orientation are shown in Fig.~ \ref{fig:fig1}(c). 

The frequencies predicted using SCUFF-EM are shown as vertical dashed lines in Fig.~\ref{fig:fig1}(d). % for a disc with $r= 4.5\mu$m and $t=400$ nm. 
The transverse modes $f_{x,y}$ and the librational modes $f_{\theta_x,\theta_y}$ respectively are mostly degenerate due to the symmetry of the disc.  Since $t<\lambda/2$ i.e. the anti-node spacing, the particle is confined to a single anti-node due to which the axial frequencies as well as the librations are much higher than the transverse frequencies.  The agreement of the translational (x-,y-, and z-) mode frequencies between the calculation and observed data is reasonable, especially  considering the disc radii can vary due to over-etching or damage during the transfer process.
In addition to the predicted motional peaks, there are two more peaks in the spectrum at approximately 8~kHz and 10.2~kHz which cannot be explained as harmonics or side-bands of the motion along other degrees of freedom. Most of the disc's trapped so far have shown at least one such additional peak in the spectrum including the spectra shown in Fig.~\ref{fig:fig3}(right). These peaks appear more prominently on the quadrant photodiode based detectors (QPD) which are positioned transverse to the trap propagation direction (see Fig.~\ref{fig:fig2}) and appear much smaller on the fiber-based detector which is most suited to collect the information of the motion along the axial direction of the trap.

\begin{figure}[b]
    \includegraphics[width=0.48\textwidth]{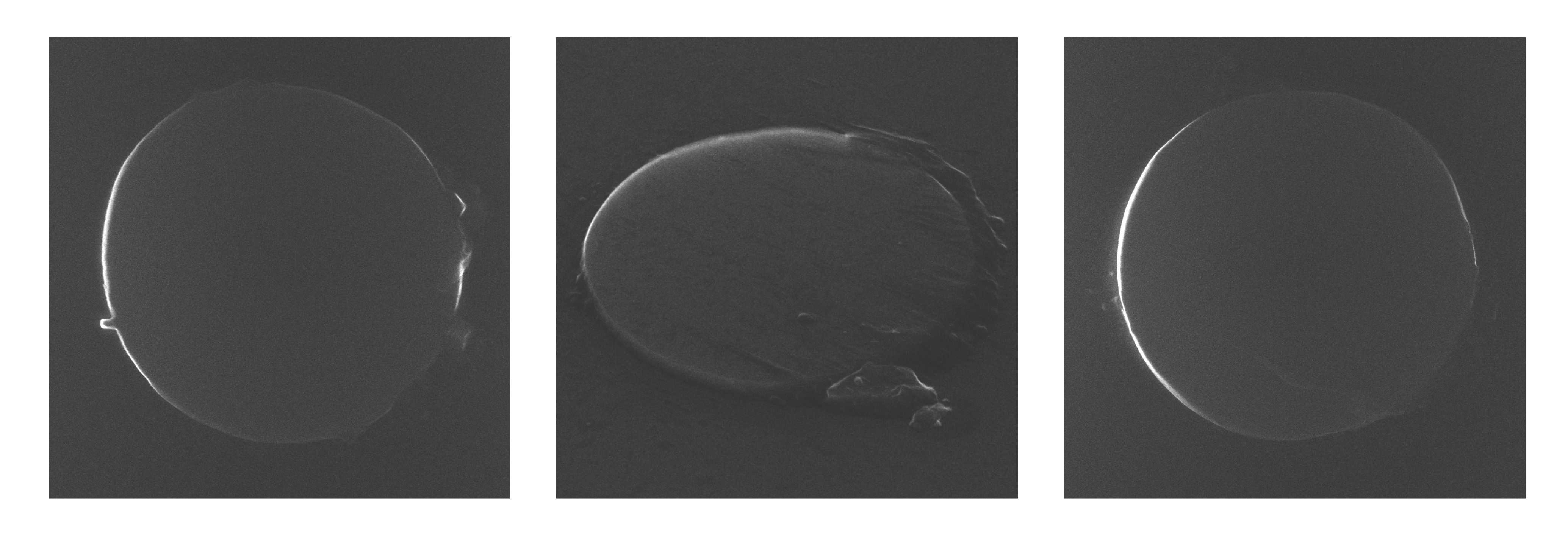}
    \caption{Additonal SEM images of the discs. The batch used for the data shown in Fig. \ref{fig:fig3} had a  diameter of (9 $\pm~1)~\mu$m. In some cases (left and middle) visible particulates have adhered to the edges of the disc for this batch which can break the cylindrical symmetry of the levitated object and lead to some variation in the trapping spectra. Viewing angle is 50 degrees with respect to the surface normal for the middle image, and zero degrees for the other two. }
    \label{fig:fig4} 
\end{figure}

We postulate that these peaks correspond to librational modes with their degeneracy broken due to particulates attached to the edge of the disc. Due to storage and handling, some of the discs exhibit dust or other adsorbed particulates that can be present near the edges of the disc, as shown in representative SEM images in Fig. \ref{fig:fig4}. Assuming an extra mass $m$ sits in the disc plane at distance $d$ from the center and
adds inertia but no restoring torque, the libration about the axis perpendicular to its lever arm then softens as $f_{\theta_y} = f_{\theta_{y0}} (1+\frac{md^2}{I_0}+\frac{2}{5}\frac{ma^2}{I_0})^{-1/2}$ and $f_{\theta_x} = f_{\theta_{x0}} (1+\frac{2}{5}\frac{ma^2}{I_0})^{-1/2}$ where $I_0$ is the single disc moment of inertia. Here for simplicity we assume that the extra mass is spherical, is oriented downwards due to the earth's gravity, and has radius $a$ and density equal to that of the silica. For example for a $1.25$ $\mu$m radius attached particle, this would change the interial mass of the disc for librations about the vertical and horizontal axes and shift the degenerate expected value of $f_{\theta_{x0}}$ and $f_{\theta_{y0}}$ modes for a bare disc from $18.5$ kHz to $18.2$ kHz and $10$ kHz respectively, and could explain the intermediate frequency peaks in Fig \ref{fig:fig3}. For example we postulate that the peak near 14 kHz in Fig. \ref{fig:fig3}(right) may correspond to libration about $\theta_y$. The peaks $f_2$ and $f_1$ in Fig. \ref{fig:fig3}(left) at $8$ kHz and $10.2$ kHz may correspond to libration around $\theta_y$ and $\theta_x$, respectively, provided there are two such attached particulates, as appears to be the case for example in the middle panel SEM image of Fig. \ref{fig:fig4}.  Future work will entail improving cleanliness in storage and handling to reduce the occurence of such adhered particulates or other defects. 

\begin{figure}[ht]
    \includegraphics[width=0.45\textwidth]{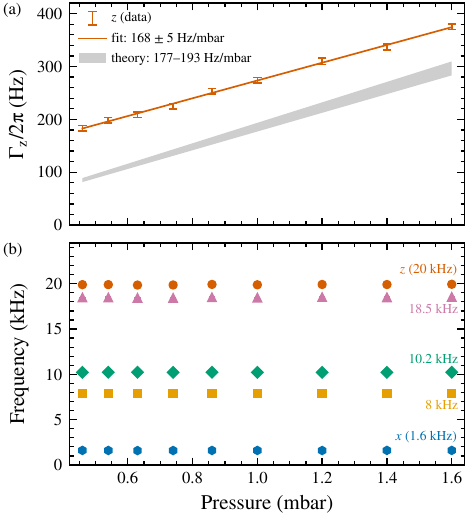}
    \caption{(a) Linewidth of the axial mode of the disc in Fig.~\ref{fig:fig1} as a function of Pressure. Each data point represents the mean of $\Gamma/(2\pi)$ obtained over five data-sets of 1-second each and the error bars show the standard error of the mean. The theoretical prediction is calculated according to Eq.~\ref{eq:Gz} with gray band representing values of $\Gamma/(2\pi)$ for all possible accommodation factors $\alpha_c\in$[0,1]. (b) Frequency vs Pressure for peaks reported in Fig.~\ref{fig:fig1}. The frequencies of the observed peaks are independent of pressure in the $0.46 -1.6$ mbar range supporting the claim that they are not associated with rotational motion.}
    \label{fig:fig5}
\end{figure}

%As a result, we performed a number of investigations to characterize the nature of these peaks and the fact that they appear larger on QPD's and much smaller on the fiber- based detector [add image of spectrum].
Since $\theta_z$ is not bound by the optical potential, the particle is free to rotate about this axis (this rotation can be thermally driven or due to optical torque arising from beam ellipticity or circularly polarized beams, etc) and if the rotation rate exceeds a threshold, the libration modes can hybridize into precession and nutation modes \cite{seberson_parametric_2019,rashid_precession_2018,rademacher_roto-translational_2026,bangFivedimensionalCoolingNonlinear2020}. Peaks corresponding to free rotation and the affiliated precession and nutation are expected to have their frequency scale linearly with pressure. To test whether $f_1$ and $f_2$ or any other peaks originate from a similar mechanism, we check the pressure scaling of all modes in the spectrum across an order of magnitude in pressure (see Fig.~\ref{fig:fig5}) and find no such scaling, providing additional evidence that these modes are purely librational or translational modes. 

%plan to add pendulum discussion back before submitting to journal

%We postulate one possible explanation for the peak at $f_1$ would be a pendular mode if there is a particulate hanging from the bottom of the trapped disc, e.g. as shown in Fig.~ \ref{fig:fig2}.  The expected frequency of such a motion in the Earth's gravitational field would be $\omega_p=\sqrt{g/l_{\mathrm{eff}}}$, where $g$ is the acceleration due to gravity and $l_{\mathrm{eff}}$ is the effective distance between the center of mass of the object and the vertical equilibrium position of the center-of-confinement of the particle in the trap. We note these could differ due to micron-sized irregular particulates affixed to the edges of the disc. For example for $l_{\mathrm{eff}}=2.5$ nm, this corresponds to a frequency of approximately $\sim 10$ kHz.   
%\textcolor{blue}{SGE: Discussion required: Extending the formula for a finite object with moment of inertia I, the pendulum formula gives $\omega=\sqrt{Mg l_{eff}/I_{pivot}}$ and when i put in numbers for our disc+ small attachment(say another tiny disc s.t. $l_{eff}$ =2.5nm, I get $\omega/2\pi \sim 10-20Hz$.Maybe I'm missing something and we can discuss this on thursday? }

%%%% AG: uncomment the following paragraph if FB cooling discussion is included...
To further characterize the motion, we also performed preliminary tests of 3-dimensional linear feedback cooling using 1064~nm light as a way to assist with mechanical mode identification. The feedback cooling light was applied from one direction of the trapping fiber using a wave-division multiplexer (WDM) as shown in Fig. \ref{fig:fig2}, as well as directed into the chamber from the side and bottom windows used for QPD imaging by use of dichroic beamsplitters. Three acousto-optic modulators (AOM) were used to independently modulate the power coming from each axis according to a phase shifted signal of the particle displacement. As an example, for the particle with spectrum shown in Fig. \ref{fig:fig3}(left) we found that we were able to cool the $20$ kHz peak most effectively
%responded best to the cooling beam 
using the cooling beam along the axial direction and the $1.6$ kHz peaks responded best to the beam along transverse direction,
%\textcolor{red}{(can we make this more quantitative?)}
further supporting that these are the normal modes corresponding to the translational degrees of freedom in the axial and transverse directions, respectively. A thorough characterization and optimization of the behavior of the discs under feedback cooling is planned for future work. 

%We did not see any conclusive response to cooling along any direction from the $f_1$ and $f_2$ peaks, leaving them unassigned.
%We used feedback cooling beams to infer which peak corresponds to which dof. 1.6kHz peak responds best to transverse cooling allowing us to infer that it corresponds to transverse (fx/fy)motion. The 20kHz peak responds best to axial cooling whereas the 7 and 8kHz only weakly respond to any cooling.  

Since we have a high-aspect ratio particle,
%\textcolor{red}{(why does high aspect ratio allow this?)}, 
we can use gas-damping rates as another method to differentiate between different motional degrees of freedom.
We can obtain the damping rate $\Gamma$ experimentally by fitting each peak in the amplitude spectrum density (ASD) using 
\begin{equation}
  S_{V}^{1/2}(f)\propto\frac{f_i^2}{\sqrt{\left(f_{i}^{2}-f^{2}\right)^{2}+f^{2}\left(\Gamma_{i}/2\pi\right)^{2}}}
  \label{eq:Svv}
\end{equation}
and extracting $\Gamma_i$ for each frequency $f_i$. We can then compare it to the theoretical formula for $\Gamma$ for each degree of freedom.
In the free-molecular regime, i.e., when the mean free path of the gas exceeds the physical dimension of the particle --- characterized by the Knudsen number $K_n>1$ ($K_n=\lambda/L$, where L is the physical dimension of the particle and $ \lambda=k_{B}T/(\sqrt{2}\,\pi d_{m}^{2}\,P) $ with $d_m$ as the diameter of the gas molecule and P as the gas pressure),  the damping rates of a rigid disk of radius $r$, thickness $t$, mass $M=\rho\pi r^{2}t$ and aspect ratio $a= r/t$  are provided in Ref. \cite{martinetzGasinducedFrictionDiffusion2018}. As an example for the axial ($z-$) motion, the theoretical damping coefficient is
%\begin{align}
%  \Gamma_{0} &= p\,\sqrt{\frac{2\pi m_{g}}{k_{B}T}}\;\frac{RH}{M}, \label{eq:G0}\\[4pt]
%  \Gamma_{x} &= \Gamma_{0}\left[\,2+\alpha_{c}\left(-\frac{1}{2}+\frac{\pi\gamma_{s}}{4}+\eta\right)\right], \label{eq:Gx}\\[4pt]
\begin{equation}
   \Gamma_{z} = \Gamma_{0}\left[\,4a+\alpha_{c}\left(1-2a+\frac{\pi\gamma_{s}}{2}\,a\right)\right], \label{eq:Gz}\\[4pt]   
\end{equation}
where $\Gamma_{0} =4P rt/(M\bar{v})$, $\bar{v}=\sqrt{8k_{B}T/(\pi m_g)}$~is the mean velocity of the gas,
%where $\Gamma_{0} = P\,\sqrt{\frac{2\pi m_{g}}{k_{B}T}}\;\frac{rt}{M}$.
% \Gamma_{\theta} &= \Gamma_{0}\,\frac{H^{2}}{3R^{2}+H^{2}}
 %   \Bigg[\,2+12\eta^{3}
%  \nonumber\\
%  &\qquad +\,\alpha_{c}\Bigg(-\frac{1}{2}+\frac{\pi\gamma_{s}}{4}+3\eta+6\eta^{2}
     %+\Big(\frac{3\pi\gamma_{s}}{2}-6\Big)\eta^{3}\Bigg)\Bigg] \label{eq:Gtheta}
%  \Gamma_{\phi} &= \Gamma_{0}\,\alpha_{c}\,(2+r). \label{eq:Gphi}
%\end{align}
%Here $\Gamma_{z}$ is the damping rates for the axial motion, $p$ is the gas pressure, $m_{g}$ the molecular mass of N$_2$, and $T$ the gas temperature. The accommodation coefficient $\alpha_{c}$ is the fraction of molecules re-emitted diffusely, and $\gamma_{s}=\sqrt{T_{s}/T}$, where $T_{s}$ is the surface temperature. 
%Here %$\Gamma_{x}$ and $\Gamma_{z}$ are the damping rates for transverse and axial motions and $\Gamma_{\theta}$  are the rates for librations, 
$P$ is the gas pressure, $m_{g}$ the molecular mass of N$_2$, and $T$ the gas temperature. The accommodation coefficient $\alpha_{c}$ is the fraction of molecules re-emitted diffusely, and $\gamma_{s}=\sqrt{T_{s}/T}$, where $T_{s}$ is the surface temperature.
% The fitted PSD linewidth (FWHM) $\Gamma/2\pi$ is obtained by fitting each peak using
%\begin{equation}
 % S_{VV}(f)\propto \frac{A}{\left(f_{i}^{2}-f^{2}\right)^{2}+f^{2}\left(\Gamma_{i}/2\pi\right)^{2}}
  %\label{eq:Svv}
%\end{equation}
%For $R=4.6~\mu$m, $L=0.4~\mu$m, $\rho=2200$~kg\,m$^{-3}$, $T=300$~K, $\alpha_{c}=1$ and $\gamma_{s}=1$, Eqs.~\eqref{eq:Gx}--\eqref{eq:Gtheta} give $\Gamma_{x}/2\pi p = 58$, $\Gamma_{z}/2\pi p = 177$, $\Gamma_{\theta}/2\pi p = 181~Hz$~Hz\,mbar$^{-1}$.% and
%$\Gamma_{\phi}/2\pi p = 57$~Hz\,mbar$^{-1}$. Varying $0.75\le\alpha_{c}\le1$ changes these by less than 4\%.
%These expressions assume a rigid, convex body, linear friction, no nearby surfaces, and a Knudsen number $\mathrm{Kn}=\lambda/R\gtrsim 9$ over our pressure range.
In Fig.~\ref{fig:fig5}(a), we find reasonable agreement between the theoretical and observed slope of damping rate versus pressure for the axial damping rate. Additional broadening mechanisms contribute to the observed vertical offset of the experimental data in Fig.~\ref{fig:fig5}(a), and are likely due to trap nonlinearities. A more thorough study of the linewidth of the mechanical resonances is planned for future work. A quantitative analysis for the other motional peaks proved to be more challenging due to the near degeneracy of the modes for the transverse vibrations and librations. Future work will involve a study over a wider pressure range to investigate using the relative damping rates as a possible method to understand the shape of the trapped object. % There is still a lot of discrepancy here though for other peaks[ref fig].  
%Finally, we note that it is possible that the discs we trapped were slightly chipped or had additional features e.g. as shown in Fig. \ref{fig:fig4} which causes the mode spectrum to be different than predicted.
We also plan to implement a polarization sensitive detection scheme such as that used in Ref. \cite{kuhn_full_2017} that will provide another means to differentiate librations from translational modes. 
%%%AG uncomment this to replace feedback paragraph if desired
%\textcolor{blue}{A thorough characterization and optimization of the behavior of the discs under feedback cooling is planned for future work. }

%whether these are librations and explain why they're much lower than predicted.
%We have noticed a variation in the size of the discs with R between 4.5-4.8 um possibily due to over-etching 

%\begin{figure}

%    \includegraphics[width=0.5\textwidth]{doublet_frequency_vs_pressure.pdf}
%\caption{\label{fig:fig4} [Preliminary Figure 5: frequency vs pressure]}
%\end{figure}
%\subsection{Theoretical Modeling}

%We are able to predict the frequencies of the translational and librational motion by performing a simulation...

%The motional spectrum for an optically trapped plate-like particle was similarly simulated in our previous work for high-aspect-ratio NaYF hexagonal prisms\cite{Winstone2022}. In that case, mode ordering was consistently predicted, but the exact values of the predicted frequencies did not agree perfectly with experimental data. This was due to the variation of the particle size within a batch which meant that the size of a particular particle in the trap could only be known approximately. For the top-down fabricated particles in this work, the size is very well known for the entire batch, so we are able to get much better agreement between predicted frequencies and experimental data.

\emph{Sensitivity to Gravitational Waves.}
For the application of gravitational wave detection, a disc-shaped particle can be suspended in a standing wave in an optical cavity. For such a configuration, as discussed in Ref. \cite{aggarwal2022searching}, the minimum detectable strain $h_{\rm{limit}}$ for a particle with center-of-mass temperature $T_\mathrm{CM}$ is approximately 
\begin{equation}
h_{\mathrm{limit}}=\frac{4}{\omega_0^2L}\sqrt{\frac{k_BT_{\mathrm{CM}}\gamma_gb}{M}\left[1+\frac{\gamma_{\mathrm{sc}}}{N_i\gamma_g}\right]}H\left(\omega_0\right), \label{eq:noise}
\end{equation}
for a particle of mass $M$ trapped in a cavity of length $L$ with response function $H\left(\omega\right) \approx \sqrt{1+4\omega^2/\kappa^2}$ 
for a cavity of linewidth $\kappa$. Here $N_i=k_BT_\mathrm{CM}/\hbar\omega_0$, \textcolor{black}{$\gamma_g=2\Gamma = \frac{32P}{\pi \bar{v} \rho t}$ is the gas damping rate at pressure $P$ with mean gas speed $\bar{v}$ for a plate or disc of thickness $t$ and density $\rho$ (assuming $\alpha_c=0$),} and $b$ is the bandwidth. The photon recoil heating rate for a plate- or disc-shaped structure is \cite{Arvanitaki:2013,Chang:2012,aggarwal2022searching}
$\gamma_{sc}=\frac{V_c\lambda\omega_0}{4L}\frac{1}{\int{dV(\epsilon-1)}}\frac{1}{\mathcal{F}_{\rm{disc}}}
$ is inversely proportional to the disc-limited finesse $\mathcal{F}_{{\rm{disc}}}$, i.e. \(2\pi\) divided by the fraction of photons scattered by the disc outside the cavity mode. The integral is performed over the extent of the suspended particle. Here $V_c$ is the cavity mode volume \cite{Arvanitaki:2013}. 
At higher pressure, the sensitivity is limited by gas collisions, while at ultra-high vacuum, $\gamma_{sc}/N_i\gamma_g > 1$ and eventually the sensitivity is limited by photon recoil heating. The thermal-noise limited sensitivity generally scales as $\sqrt{P}$ in the gas-dominated, ballistic regime.  At a given pressure, we can adopt the figure of merit defined in Ref. \cite{winstone_optical_2022}, $\eta_{\mathrm{thermal}}$ which describes the dependence of $h_{\rm{limit}}$ on particle mass, geometry and trapping frequency in the thermal noise limited regime as
$\eta_{\mathrm{thermal}} \equiv \omega_0^2\sqrt{M \rho l}$, where $\omega_0$ is the trapping frequency along the cavity axis, $l=t$ for plates of thickness $t$, respectively.  
In this work, e.g. for the particle shown in the right panel of Fig. \ref{fig:fig3} with $\omega_0/(2 \pi) \approx 35$ kHz, $\eta_{\mathrm{thermal}} = 340$ Hz$^2$kg/m, representing an advance by approximately a factor of 3 over previous work.  Due to the larger mass, significant improvements in the figure of merit for the corresponding recoil-dominated regime $\eta_{\mathrm{recoil}} \equiv \omega_0^{3/2}\sqrt{M/\gamma_{sc}}$ estimated in Ref. \cite{winstone_optical_2022} may also be achievable depending on the disc-limited cavity finesse and cavity mode volume used in a cavity based GW detector. %Correspondingly, we define figure of merit $\eta_{\mathrm{recoil}} = \omega_0^{3/2}\sqrt{M/\gamma_{sc}}$ for sensitivity to GWs in the photon-recoil dominated limit. Assuming a disc limited finesse of order $10^4$, we estimate $\eta_{\mathrm{recoil}} = ...$ and 
%Will we include the same equations and figures of merit as we did in the hexagon paper??

%\section{Preliminary Feedback Cooling}

%We have discussed possibly showing some preliminary cooling results...

\emph{Summary \& Outlook.}
We have established a high-efficiency method to load small quantities of top-down lithographically nanofabricated high-aspect-ratio silicon dioxide discs into an optical standing wave trap in vacuum. We expect this approach to allow for trapping of customized nanofabricated particles which are well suited for the application of high frequency gravitational wave detection when suspended in an optical cavity. We have reported the 
%displacement 
amplitude spectral density observed for such trapped disc-like particles and have shown reasonable agreement with theoretical expectations for the observed trapping frequencies and the scaling of the damping rate with pressure for the axial mode, which is of greatest interest for the future application of gravitational wave detection. Refinement of particle handling methods to minimize impurities or particulates on the trapped discs may also be beneficial for improved understanding of the particle's motional dynamics.  Future work will involve 
%the implementation 
further optimization of feedback cooling to better understand dynamics of the translational and librational motion and to study prospects for trapping in the ultra-high vacuum regime,  enabling ultrasensitive detection of forces and acceleration.

\begin{acknowledgments}

%We wish to acknowledge the support of \dots.

%\textcolor{blue}{SG: Confirm facilities used.}
%\textcolor{green}{MN: What work was done at NUANCE?}

This work made use of Northwestern University's NUANCE Center and its NUFAB facility (RRID: SCR\_017779), which has received support from the IIN and Northwestern's MRSEC program (NSF DMR-2308691).
This work was also carried out at the Pritzker Nanofabrication Facility, part of the Pritzker School of Molecular Engineering at the University of Chicago, which receives support from Soft and Hybrid Nanotechnology Experimental (SHyNE) Resource (NSF ECCS-2025633), a node of the National Science Foundation’s National Nanotechnology Coordinated Infrastructure {[RRID: SCR\_022955]}. 
This research was supported in part through the computational resources and staff contributions provided for the Quest high performance computing facility at Northwestern University which is jointly supported by the Office of the Provost, the Office for Research, and Northwestern University Information Technology.

The authors would like to thank Andrew Dana for his contributions toward loading the discs into the pipette launcher. We would like to acknowledge discussions with Cris Montoya, Julia Krueger, Joseph Chamberlain, and Daniel Jacobson.

S.G.E is supported by CIERA Board of Visitors Graduate Fellowship. \textcolor{black}{A.L. acknowledges support from the Fannie and John Hertz Foundation through a Hertz Fellowship. N.A. is supported by the UC Noyce Initiative and the Department of Energy award number DE-SC0009999.}
AG is supported from NSF grants PHY-2409472 and PHY-2111544, DARPA, the W.M. Keck Foundation.
AG is also supported by the ``Table-top experiments for fundamental physics" program, sponsored by the Gordon and Betty Moore Foundation, Simons Foundation, Alfred P. Sloan Foundation, and John Templeton Foundation. AG is funded in part by the Gordon and Betty Moore Foundation through Grant GBMF12328, DOI 10.37807/GBMF12328. The work of AG is partially supported by the Alfred P. Sloan Foundation under Grant No. G-2023-21130. %\textcolor{red}{Moore Foundation and Sloan Foundation repeated twice?}

\end{acknowledgments}

%\nocite{*}

\bibliography{apssamp}% Produces the bibliography via BibTeX.

\end{document}